\documentclass[sigconf,screen]{acmart}

\usepackage{enumitem}
\usepackage{framed}
\usepackage[most]{tcolorbox}
\usepackage{tikz}
\usetikzlibrary{positioning, arrows.meta}
\usepackage{hyperref}
\AtBeginDocument{%
  }

\setcopyright{cc}
\copyrightyear{2026}
\acmYear{2026}
\acmDOI{XXXXXXX.XXXXXXX}
\acmConference[Conference'26]{Make sure to enter the correct
  conference title from your rights confirmation email}{June,
  2026}{UK}
\acmISBN{978-1-4503-XXXX-X/2018/06}
\begin{document}


\title{Children Are Not the Enemy: Child-Fit Security as an Alternative to Bans and Surveillance}

\author{Kopo M. Ramokapane}
\email{marvin.ramokapane@bristol.ac.uk}
\orcid{0000-0001-8420-3929}
\affiliation{%
  \institution{University of Bristol}
  \country{United Kingdom}
}

\author{Rui Huan}
\email{rui.huan@bristol.ac.uk}
\orcid{0000-0002-8713-6458}
\affiliation{%
  \institution{University of Bristol}
  \country{United Kingdom}}

\author{Zaina Dkaidek}
\email{zd420@bath.ac.uk}
\orcid{0009-0000-6865-4201}
\affiliation{%
  \institution{University of Bath/Bristol}
  \country{United Kingdom}
}

\author{Awais Rashid}
\email{awais.rashid@bristol.ac.uk}
\orcid{0000-0002-0109-1341}
\affiliation{%
  \institution{University of Bristol}
  \country{United Kingdom}
}

\renewcommand{\shortauthors}{Ramokapane et al.}

\begin{abstract}
Digital technologies are now central to children’s learning, play, communication, identity formation, and social participation. Yet dominant approaches to children’s online safety often rely on containment mechanisms, including bans, age gates, parental controls, monitoring, and screen-time restrictions. These approaches can be useful in specific contexts, but they often frame child protection primarily as a problem of restricting access to systems designed for adults. In this paper, we argue that this framing is inadequate for children’s digital lives and insufficient as a security paradigm. We propose Child-fit security, a design paradigm in which technologies likely to be used by children treat a child as legitimate users, not attackers to be excluded, vulnerabilities to be patched, or risks to be managed. In this paradigm, children’s wellbeing, development, privacy, safety, agency, and rights become core security requirements. This shifts the focus of protection from apps, accounts, and data to the \emph{child–system} relationship, which means protecting both the child and their participation. We conceptualise child-fit security, contrast it with containment-oriented approaches, define its core principles, and discuss its implications for security design. We conclude by presenting a research agenda for making child-fit security operational.
\end{abstract}



\received{20 February 2007}
\received[revised]{12 March 2009}
\received[accepted]{5 June 2009}

\maketitle

\section{Introduction}
Children are not the enemy, poorly designed systems are. Yet, when technology fails to protect them, they end up bearing the consequences (e.g., bans and restrictions) while the systems that failed them remain unchanged. Throughout the last decade, every time a system fails a child or young person, governments and societies respond not by fixing the vulnerability, but by restricting the user who in this case is least equipped to advocate for themselves. While the security community has learned, at least in principle, that ``users are not the enemy''~\cite{adams1999users}, protective efforts continue to treat children as the vulnerability rather than interrogating the systems that expose them to harm and holding them accountable for designing safer environments.

Technology has become part of everyday life and children are no longer occasional users, they grow up with it~\cite{wang202312}. In the US, 96\% of teenagers use the internet every day while 45\% report being online constantly~\cite{pewTeensInternetDeviceAccess2025}. In the UK, Ofcom similarly reports that 96\% of 8- to 14-year-olds used YouTube and 95\% used Google Search~\cite{ofcomChildrenMediaUseAttitudes2025,ofcomChildrenPassiveOnlineMeasurement2025}. Also, Common Sense Media found out that 40\% of children have a tablet by age two, and nearly one in four have a personal phone by the age of eight~\cite{mannCommonSenseCensus2025}. These statistics suggest that children are entangled in technology and it plays a huge part of their lives. Nowadays childhood unfolds with and through technology; children play, learn, socialise and communicate through various technologies, making technology inseparable part of their cognitive and social development. 

Yet, the safety and security of technologies and platforms used by children remain contested. Public debates about how they should be protected are often shaped by stories of severe harm. Basant Khaled was a 17-year-old Egyptian schoolgirl from Gharbia governorate who died by suicide in 2022 after being targeted a young man she rejected~\cite{bbc2022basant,egyptianstreets2022basant}. She was allegedly sent a deceptive link, after which personal photographs from her phone were accessed and manipulated into compromising images. These images were then allegedly used to blackmail her, including threats that they would be shown to her family and wider community. 
Such cases make the need for child protection online appear obvious and urgent, and rightly so. However, they also shape how child online safety is imagined and defined; children are vulnerable, online environments are dangerous, and therefore protection means removing them from such risk spaces.

This framing is not new. Historically, children have been treated as incidental users of adults focused systems, their needs have been addressed reactively than proactively and mostly through policy interventions or denial of access altogether~\cite{prendergast2026youth}. When cases like Basant occur, the natural response has been restrictions, children are denied access, age verification is imposed, surveillance increases, or children are removed from spaces deemed risky. Australia's under-16 social media restrictions, which came into effect on 10 December 2025, illustrate this broader turn toward age-based exclusion~\cite{bbcAustraliaSocialMediaBan2025}. Though these measures are not ill-intended, they often misidentify the source of the problem. Risk is associated with the presence, access or behaviour of the child, rather than the design, governance, and commercialisation of the systems they engage with. It is often forgotten that children do not choose what they see, encounter, or engage with, but it is through algorithms, platform incentives, interface design, data practices, and weak protective mechanisms and architectures~\cite{volz2026age, prendergast2026youth, archer2025coming}. This represents a significant gap between the lived digital experience of children, their protection and the design of systems they use daily.  

In this paper, we argue that harm-based cases should not be the only basis on which child online protection is conceptualised and understood. The case of \emph{Mats Steen} offers a different perspective. Mats was a Norwegian man who had lived with Duchenne muscular dystrophy since childhood~\cite{king-schreifels2024ibelin,schaubert2019mats}. As his condition deteriorated and increasingly restricted his physical mobility, much of his social interaction took place online. Through \textit{World of Warcraft}\footnote{Note: We are not in anyway suggesting that children have not been harmed in World of Warcraft.}, where he was known as \emph{Ibelin}, he formed friendships, participated in a community, and developed a social identity that his physical circumstances made difficult to sustain offline. When he died in 2014, his parents later learned that his online life had involved sustained relationships, emotional support, and a strong sense of belonging. 

These two cases are not outliers, but they illustrate a crucial tension in child online protection. When protection is imagined primarily through cases such as Basant’s, policy and design responses tend to prefer restriction, surveillance, and exclusion~\cite{feal2020angel,archer2025coming,prendergast2026youth}. However, when protection also takes seriously cases such as Mats’s, such responses appear incomplete and potentially harmful. Restrictions may reduce exposure to certain risks, but they may also produce other harms and cut children off from spaces of connection, identity, care, and participation. For this reason, the challenge is not only that the current paradigm define protection too narrowly as the prevention of direct harm. It is also that it often positions children themselves as the risk to be managed; users who must be monitored, verified, or excluded~\cite{wang2021protection,volz2026age, prendergast2026youth}. This framing obscures the more fundamental issue: children are already part of digital society, yet many digital systems are not designed with their rights, vulnerabilities, and forms of participation in mind. We believe that a more adequate approach to child online protection must therefore shift the site of intervention from the child to the system (child-system relationship). Protection should not depend primarily on removing children from digital spaces, but on transforming those spaces so that children can participate safely, meaningfully, and with dignity.

Consequently, we propose a new frame of thought around child protection: \textbf{\emph{Child-fit security}} paradigm. Child-fit security is a security design paradigm in which technologies likely to be used by children treat children’s wellbeing, development, privacy, safety, agency, and rights as core security requirements rather than afterthoughts. The key shift in this paradigm is that we are not only ``securing'' the app, account, data, or network that the child’s use of technology engages with; we are also securing the relationship with the system. Current approaches tend to focus on the former while neglecting the latter. As a result, they often emphasise exclusion, surveillance, and restriction. 

When children are not totally excluded from a system, responsibility for their safety is often shifted down to them and their parents. They are expected to manage risks through behaviour, monitoring, settings, or withdrawal of access, regardless of their skills, knowledge, or capacity to understand the system~\cite{cranor2014parents,akter2022parental}. While some restrictions may be justified, there is often less scrutiny of the system itself or of the design choices that shape children’s experiences. Harms end up being framed as failures of children’s behaviour or parental supervision, rather than as foreseeable consequences of systems not designed with children as legitimate users.

Child-fit security brings the child to the centre of security design. It argues that children should be considered throughout the development of systems they are \emph{likely} to use. Protection should be considered from the start, not as an afterthought or something bolted onto systems after they have already been designed for adults. Likelihood is important here. We argue that, for every system being designed, developers should consider whether children are likely to use it. If they are, then children should be considered in its design. The premise is that children are users, not mini-adults, and their evolving capacities, developmental differences, relational needs, and specific rights should be recognised in system design~\cite{wang2021protection}. This does not mean every system should be made available to children. Some services are intended for adults, and such systems should be designed so that they are not accessible to children. However, systems that may reasonably be used by children should not simply exclude them, surveil them, or retrofit protections around adult-oriented designs. With this in mind, this paper develops child-fit security as both a critique of current approaches and a constructive paradigm for security design. We make the following contributions: 
\begin{itemize}
    \item We propose a new paradigm for child online protection: child-fit security. This paradigm argues against blanket bans and surveillance for children using systems in which they are part of the user group. It argues that children’s evolving capacities, developmental differences, relational needs, and specific rights should be first-class design considerations.
    \item We critique the existing paradigm, showing how well-intentioned approaches can lead to unintended harms.
    \item We conceptualise child-fit security by defining its core principles and discussing its implications for security design. Building on this, we propose a research agenda to operationalise child-fit security.
\end{itemize}

\section{The Current Paradigm: Online Child Safety as Containment}

Protecting children did not start with smartphones, it dates back to the years when the internet became a public place. In 1996, the United States enacted the Communications Decency Act as part of the Telecommunications Act of 1996, marking one of the first major government attempts to regulate online content, particularly sexually explicit material accessible to minors, in the name of child protection~\cite{zeigler2023cda230}. As a result, it established how many governments and platforms would approach child protection. The Children’s Online Privacy Protection Act (COPPA) of 1998 represented a significant early effort to regulate children’s online privacy by requiring digital services to obtain verifiable parental consent before collecting personal information from children under the age of thirteen. However, this move contributed to the normalisation of thirteen as a regulatory threshold for digital participation. In practice, many platforms responded not by developing child-centred safety architectures, but they excluded under-thirteen users and relied totally on self-reported age gates. This shifted responsibility away from platform design and toward the child’s act of misrepresentation. For instance, if a child lied about their age, this was classified as circumvention rather than the platform having inadequate safeguards. This resulted in a model in where age ages were seen as child protection.

This regulatory pattern became a norm and was popularised by the rise of social media platforms, from myspace to TikTok. During the Myspace era, efforts largely focused on predatory behaviour and child exploitation, and it was again addressed through policy changes and congress hearings until the platform collapsed~\cite{marwick2008catch,house2006dopa}. However, when Facebook came, it grew very fast and regulation could not keep up but children continued to face harm. Then, Instagram, TikTok and other platforms with the same engagement-driven model (e.g., attention economy) came and every time a harm was identified the questions have never been about the algorithms, notifications, attention-maximising interfaces or data, but about the age of the user. This was also evident when Frances Haugen~\footnote{https://www.franceshaugen.com/} leaked internal Meta research concerning Instagram’s effects on teenage users~\cite{ukparliament2021draftonlinesafetybill}. Meta's responded by announcing tools for parents over their children’s accounts and stricter age verification. The move treated children as unauthorised and passive users who had no say in interventions, while the report clearly showed that the platform worsened body image issues, heightened negative feelings, and, in some cases, was linked to increases in suicidal ideation~\cite{BARRY20171,nasem2024introduction}. Despite all this, the questions did not challenge the way these systems were designed or how child protection is conceptualised.

This paradigm continues to be a standard, and governments around the world are now promoting this way of thinking through law. Australia has moved to ban children under 16 from social media. Some US states and countries have passed or are proposing similar restrictions. The UK has enforced and intensified age-verification requirements. All these measures suggest that the problem is the child and that children should stay away from technology. We need a paradigm that challenges these ideas and recognises children as legitimate users whose participation needs to be protected rather than denied.

\subsection{Existing Interventions to Child Protection}
Over the years, several interventions have been proposed to protect children from digital harms. These have included both technical and policy measures. Technical measures include client-side scanning, parental-control software, device-level restrictions, school filtering systems, and platform age-assurance tools. Policy and institutional measures include school internet policies, social media age bans, and phone-free school initiatives such as Yondr pouches. While these approaches differ in scale and method, they all assume and treat a child as a primary risk to be managed, users to be monitored, or subjects to be excluded, rather than a legitimate user in digital environments. 

\subsubsection{Client Side Scanning}
In the early 2020s, Apple proposed a client-side scanning solution to detect known child sexual abuse materials on users’ devices they can be uploaded to iCloud. While Apple argued that this was targeting abusive material (which is a legitimate safeguarding goal), many security researchers have argued that client-side scanning could undermine the security properties of end-to-end encryption, introduce new attack surfaces, normalise device-level inspection, and create infrastructures vulnerable to function creep~\cite{abelson2024bugs,struppek2022learning}. Others said this could be repurposed for other monitoring goals~\cite{org_uk_open_letter_2023, eu_csar_open_letter_2023,edri_eu_countries_no_csar_2023}. In this model, the child was positioned as a subject to protect within detection pipelines, while safety and privacy were presented as mutually exclusive objectives.

\subsubsection{Age-based Access Controls (Social Media Bans)}
Another dominant approach is exclusion through age-based access control. In this model, social media bans, age gates, and age assurance systems are used to reduce exposure to harm by keeping children outside certain systems and platforms~\cite{reuters2026_social_media_restrictions}. In Dec 2025, Australia introduced the world’s first outright ban on social media for under-16s, which prevents them from accessing major social media platforms including TikTok, X, Facebook, YouTube, Instagram, and Snapchat~\cite{oaic_social_media_minimum_age_2025,esafety_age_restricted_platforms_2026,esafety_social_media_age_restrictions_2026}. Similarly, there have been policy debates in the UK and other countries considering whether children’s social media use should be reduced through age-based restrictions. In Brazil, rather than simply banning social media access, the Digital Statute of Children and Adolescents (ECA Digital), in force from March 2026 requires social media accounts for under-16s to be linked to legal guardians~\cite{baker_mckenzie_digital_eca_2026}. Technology companies must also implement age verification to block under-18s from inappropriate content~\cite{demarest_eca_digital_2026}. China provides a broader restriction at a national level. In mainland, major foreign social media platforms such as Facebook, X, YouTube, and Instagram are blocked through the country’s internet censorship system~\cite{freedom_house_china_fotn_2024}. Since 2021, under-18s have been limited to playing online games only during defined time windows, and minors can only access online games if they have registered with real names and completed the verification~\cite{china_state_council_online_gaming_2021}. Though these measures may be politically attractive because they offer a clear regulatory boundary, they shift the responsibility from platforms to the presence of a child. While this model may reduce certain harms, they treat children as unauthorised users rather than addressing reasons that make digital systems unsafe for them. 

\subsubsection{School-based Restrictions}
School-based restrictions have become an important field of child protection because many children spend a large part of their day in school. These interventions mainly include two ways: devices ban in the classroom and platform blocking. UNESCO reports that \emph{phone bans in schools} are no longer isolated national practices but spreading worldwide~\cite{unesco_phone_bans_2026}. Phones are required to be left at home, placed in lockers, or locked away for the full school day, as reflected in school rules and policy in the Netherlands, France, and Chile~\cite{netherlands_school_phone_rules_2025,france_school_phone_rules_2024,chile_mobile_devices_schools_2026}. Some schools also adopt Yondr pouches\footnote{\url{https://www.overyondr.com/phone-free-schools}} in schools, where students are required to place their devices in locked pouches during the school day, preventing access while still allowing them to keep the devices in their possession. 
Schools also increasingly apply \emph{platform blocking} on school-managed devices. In practice, they often rely on broad deny-all rules that block entire services such as YouTube or GitHub rather than specific content~\cite{greenwich_youtube_block_2025, greenville_youtube_block_2023,guardian2026_la_youtube_block}.

While these practices are often justified by concentration, wellbeing, or safeguarding, schools rely on removing or blocking devices or platforms altogether rather than consider how technology can be integrated into school life. This may overlook the fact that technology is not always used for entertainment or risks such as cyberbullying; for young people, it can also be as an accessible way with school work and problem-solving. Teachers may ask students to use GitHub for computing classwork, but students then have to do it at home (outside school) due to school firewall rules, where it is less unsupervised and may create additional risks. In this model, children are positioned as subjects to be managed rather than as users whose needs should help create a safe digital environment.

\subsubsection{Filtering and Monitoring Technologies}
Child protection is also enacted through ongoing monitoring, filtering, and management technologies, not only through bans or restriction. In some cases, systems are underpinned by digital identity and age verifications mechanisms, as a gateway for classifying users and then enforcing different controls. These interventions allow adults to monitor and manage what children can access, who they can interact with, and what activities are permitted~\cite{feal2020angel}. At the \emph{household-based} level, key tools such as Google Family Link\footnote{\url{https://families.google/familylink/}}, Apple Screen Time\footnote{\url{https://support.apple.com/en-gb/105121}}, Nintendo Switch parental controls\footnote{\url{https://www.nintendo.com/en-gb/Support/Parental-Controls/Supervise-your-child-s-gameplay-1197305.html}}, and Amazon Kids+\footnote{\url{https://www.amazon.com/ftu/home}} allow parents and guardians to set screen-time limits, filter content, restrict app downloads and usages, and manage communication features. At the \emph{platform-based} level, some gaming platforms  have their own in-platform controls, for example, Roblox~\cite{roblox_parental_controls} embed parental controls for children under 13, by requiring age verification via government-issued ID or credit card to establish linked accounts. Once verified, parents can configure content maturity settings, monitor spending, and limit interactions to ``trusted-friend'' only. Commercial safety products, such as Net Nanny\footnote{\url{https://www.netnanny.com/}}, adopt similar features by providing cross-platform content filtering and real-time activity alerts. At the \emph{school-based} level, safety technologies are frequently deployed to monitor student activity on school-managed systems~\cite{gaggle_school_surveillance_2025,the74_ohio_gaggle_2024, the74_survey_goguardian_2022,loilo_web_filter}. For example, Japan's LoiLoNote allows teachers to selectively permit access to specific URLs or educational videos~\cite{loilo_web_filter}. While these systems are applied across diverse contexts, child protection is pursued through monitoring children's activity, narrowing their choices, and giving adults greater control over access and interaction. In this model, these technologies often prioritise top-down control and surveillance over children's participation in designing systems that reflect their own needs and agency.

\subsection{Unintended Consequences of Existing Approaches to Child Protection}
Existing approaches to child online protection can produce unintended consequences when they rely mainly on surveillance, exclusion, filtering, or restriction. While these measures may reduce some exposure to harm, they can also weaken security, increase data collection, force children into less visible online spaces, restrict legitimate access, and undermine trust between children and adults. In this section, we discuss various consequences that may result because of these interventions. 

\subsubsection{Weakening Security for Everyone}
Some child online protection interventions may unintentionally weaken the security systems that also protect children. Security experts and digital rights groups have argued that such measures can undermine the guarantees of end-to-end encryption by creating scanning or access mechanisms that may later be expanded, misused, or exploited~\cite{internetsociety_2023_clientsidescanning,abelson2024bugs}. This is illustrated by recent UK debates over encrypted services. In 2021, the UK government supported the development of ``safety tech'' tools intended to detect child sexual abuse material in encrypted environments~\cite{dcms_homeoffice_2021_safetytech,peersman2022CSAM}. In 2025, Apple withdrew its Advanced Data Protection feature for UK users after the UK government reportedly wanted access to encrypted iCloud data. Apple now states that Advanced Data Protection is no longer available in the UK~\cite{apple_2025_adpuk,reuters_2025_appleadpuk}.

However, children use encryption or private communications for legitimate and protective purposes such as help-seeking, support, family contact, identity exploration, and protection from abusers. UN Children’s rights frameworks also recognise that privacy and protection from harm must both be safeguarded in the digital environment~\cite{unicef_2020_encryptionchildren}. Moreover, even if scanning occurs before content is encrypted or after it is decrypted, this undermines the security promise of end-to-end encryption which could lead to making private devices part of monitoring systems~\cite{internetsociety_2023_clientsidescanning}. The unintended consequence is that child protection becomes linked to weakening the protection of systems or normalisation of surveillance, while their other rights such as privacy may be compromised~\cite{rashid2025weakening}.

\subsubsection{Inequality and Exclusion}
Restriction can create or widen existing inequality. Approaches such as blocking, banning, and phone-free schooling may reduce immediate exposure to harm, but they can also reduce opportunities for learning, socialising, playing and communicating. Prior research has shown that access to the internet is closely connected to opportunities for learning, participation, information-seeking, creativity, and social connection~\cite{notley2009young}.

A child from a wealthy family in a connected environment has alternatives such as tutors, educational tools, access to supervised devices, and social networks built around school other forms of engagement. However, a child from a rural place, a Global South region, or low-socioeconomic household may have limited or no access to these alternatives. For that child being denied access means no help for homework, no peer connection and access to information (e.g., creative tools, developer materials, health information, suicide-prevention material). Unequal digital access has consistently been linked to wider educational and social inequalities~\cite{helsper2021digitalinequalities,unicef2025childhooddigitalworld}. For instance, New America argues that some filters prevent LGBTQ youth from accessing vital online resources, including health information and crisis support~\cite{newamerica2023antilgbtqfilters}. In such cases, filtering may remove access not only to risky content but also to protective or developmental resources. 

While age assurance is often framed as privacy-preserving, it may also introduce equity concerns. They may misclassify people and even be burdensome for many users. Children without formal identity documents, sharing devices, children with disabilities, trans and gender-nonconforming children, and children from marginalised communities may be disproportionately affected by inaccurate or burdensome verification systems. In UK such systems are required to use approaches such as document verification and facial age estimation to ensure age is accurate and reliable~\cite{ofcom2025agechecks}.

Some approaches may also unintentionally reduce children’s chances for guided learning. If children are protected primarily by removal, they may have fewer chances to develop practical digital skills such as recognising manipulation, managing privacy settings, responding to unwanted contact, reporting abuse, or negotiating peer conflict online~\cite{crc_gc25}.

\subsubsection{Displacement of Accountability and Responsibility}
Many existing approaches (e.g., bans and parental-control platforms) shift responsibility away from platforms and system designers onto children, parents, and schools. Parents are expected to install and understand controls, schools are expected to filter and regulate the use of devices~\cite{cranor2014parents,akter2022parental}, while children are expected to comply with restrictions. In some cases, general users may be required to prove their age. 

In countries where restrictions and bans are in place, the platforms may feel no need to improve the safety of their systems on the assumption that children are not present. This weakness the whole idea of safety. Similar to device bans and phone-locking systems, policies created in the ``best interests of the child'' can become paternalistic if children’s own perspectives are excluded. Research (e.g.,~\cite{hartikainen2019childrensdesign,unicef2025bestinterestsdigital}) stresses that children’s voices are often absent from decisions about what is in their best interests in digital environments. The unintended consequence is that child safety becomes something done to children rather than designed with them. 

There is also possibility of parents shifting from care and guidance to relying on continuous monitoring. For instance, schools and parents may not invest in teaching children about online safety but just rely on monitoring what they do on their devices. Interventions and designs that exclude children or their own experiences may also lead to misidentifying assets, adversaries, harms, and acceptable trade-offs which are important to consider to effectively protect children.

\subsubsection{Privacy Invasive Identity Verification Systems}
While age assurance may have a role in applying age-appropriate protections, it can create new privacy risks when implemented as an exclusion mechanism. To determine whether a user is a child, platforms may be required to collect data that has traditionally been optional or unnecessary for access. This may include identity documents, facial images, biometric signals, behavioural inference, or verification through third-party providers~\cite{jarvie2024onlineageverification,shaffique2025behaviouralprofiling}. As a result, this approach does not only collect data from children but from all users.

These systems may also create new capabilities for classifying, profiling, or ranking users by age. This is concerning because these systems may rely on algorithmic analysis of users’ online activities to infer age, resulting in risks for both children and adults. Shaffique argues that behavioural profiling for age assurance may affect the rights of online users~\cite{shaffique2025behaviouralprofiling}. Moreover, such system may produce false positives locking genuine users out since they rely on probabilistic outcomes. False positives may wrongly classify adults as children, while false negatives may allow children to access spaces intended for adults. This is concerning because if a genuine user is locked out, they may not assess lawful content or be forced to disclose more personal information in order to be correctly classified.

\section{Child Protection is a Security Problem}
In this section, we aim to remind the security community that online child protection is not only a policy problem but a security problem. 
 
While there has been rapid policy response toward online child safety obligations, mainstream security research has not yet matched the speed at which policy is responding in this area. As a result, technical research agenda in this area remains undeveloped. Existing work is scattered across usable privacy, parental controls, age assurance, educational apps, and youth security literacy, and these efforts are often published in HCI and policy venues. This gap is critical because now the space has been established through regulation, compliance, and governance debates. The limited (or lack of) engagement from the security community has left the area without strong security framing; the problem space has been defined through obligations not security properties.

Because of this, we believe the space has now been regulated before it could be technically guaranteed. Platforms are encouraged to deploy various technical measures, yet their protection capabilities remain immature. The field still lacks sufficient evidence about what works, what could be circumvented, their usability by children and parents, and whether these measures meaningfully reduce harm. A platform can appear compliant because it has an age gate or certain settings while the underlying infrastructure fails to guarantee the promises. Moreover, platforms may remain less innovative because they may have little incentive to go beyond what is required.  

In this paper, we argue that online child protection is a security problem because it involves protected assets, adversarial actors, system boundaries, design choices, technical controls, misuse cases, failure modes, and assurance requirements specific to children. When it comes to children, the assets to protect go beyond data, account integrity or service availability, they include children’s attention, privacy, agency, developmental space, bodily and emotional safety, meaningful choices, and freedom from manipulation and coercive design. The relevant threats also extend beyond hackers but may include predators, malicious peers, abusive carers, and platform designs that expose children to risks they cannot reasonably understand or avoid.

We argue that a harm becomes a ``security problem'' when it is produced, amplified, or mitigated by technical systems that define access, visibility, identity, recommendation, data collection, contact, and control. As a results, when a system that a child may use fails, the failure is not only a policy failure but a security failure. In other words, harms are not just bad outcomes, they are security failures of the system. For instance, \emph{compulsive use} is a failure of protecting a child's attention. This means a system has failed to protect a child's \emph{ability to disengage} which may have resulted through \textit{notifications, streaks, recommendations}. The security community should therefore treat online child protection as part of its core agenda. There is a need for frame of thought that asks what must be protected, from whom or what, through which controls, under what assumptions, and with what evidence that those controls work.  

\section{New Paradigm: Child-Fit Security}

\begin{framed}

    \textit{``We do not make playgrounds safe by banning children from climbing and running around. We make playgrounds with soft landings, appropriate heights, visibility, challenging, and room for growth.''}

\end{framed}

\begin{figure} [ht]
    \centering
    \includegraphics[width=0.65\linewidth]{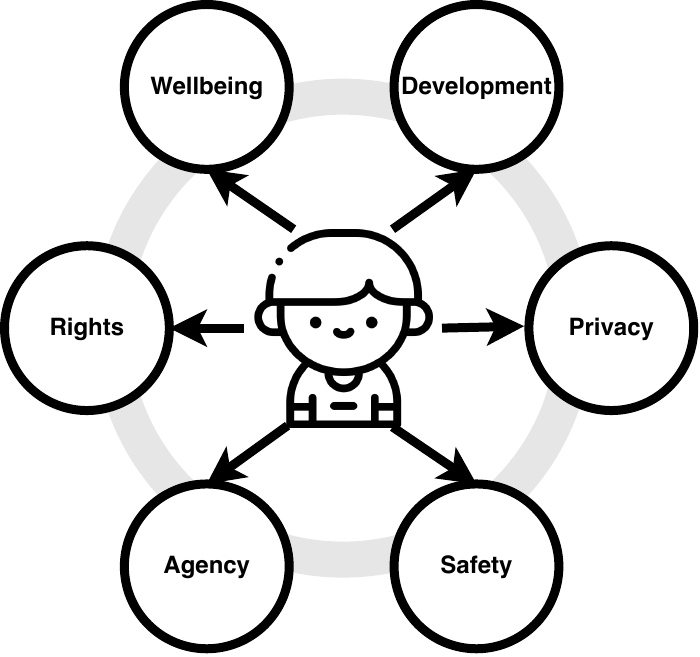}
    \caption{\emph{Child-Fit Security} core requirements.}
    \label{fig:core-requirements}
    \Description{Core requirements of Child-Fit security. Rights, Wellbeing, Development, Privacy, Safety, Agency}
\end{figure}

\textbf{Child-fit Security} is the design and governance of digital systems in which children's developmental needs, rights, privacy, safety, agency, and wellbeing are modelled as first-order security requirements. 

This paradigm frames child protection as a design problem and children as legitimate users with distinction and evolving capabilities, vulnerabilities and rights. It shifts us from the idea that children are edge cases, and their protection is a matter of parental controls, content moderation, and regulatory compliance. Existing security paradigms were grounded in confidentiality, integrity and availability where users are adults operating systems in contexts where they are assumed to be mature and legally capable for making informed decisions. Child-fit security recognises that these assumptions do not hold any more, adults are not the only users and children use these systems as well. Moreover, children are not always mature enough to make informed decisions about their safety. A child user is \emph{a developing individual who engages with digital systems for meaningful purposes and whose interaction with those systems can affect their privacy, identity, relationships, learning, autonomy, safety, and wellbeing (Figure~\ref{fig:core-requirements})}. However, current efforts leave a significant gap in how children protection is designed and the realities of modern childhood. They often do not consider these different aspects.

\begin{framed}
    \textit{``When cooking for children, we do not only think about just filling their stomachs; we consider their nutrition, development, capabilities, and interests. A steak may be nutritious, but not suitable for a baby; milk may be appropriate for an infant, but not sufficient for an older child. Same with utensils, we may give an infant a bottle, a young child a spoon, and an older child a fork, not because one tool is universally safer than another, but because each fits a different stage of development.''}
\end{framed}

This new paradigm emphasises protections that are fitting to children. The notion of ``child-fit'' comes from the principle that systems and security should be made to fit their users rather than the user fitting the system. Current systems are not designed to fit children’s interactions (or lack thereof) and their developmental needs. Consequently, their interaction and development needs remain outside the designers’ threat models leaving them excluded or vulnerable in systems they have access to. This new paradigm argues that to effectively protect or “fit systems” to children, we first need to acknowledge and recognise children as legitimate users, then conceptualise what security means to and for them. 

This frame is important because security is often designed based on how the user is imagined. If children are not imagined as legitimate users, then the risks they face are likely to be misunderstood, minimised or ignored. Recognising children as users makes their forms of insecurity visible. It provides the foundation for accurate threat modelling, appropriate security requirements, and system design that reflect the realities of childhood and their vulnerabilities. It also allows us to examine children through security perspectives that are often overlooked; children as active participants whose needs should be central rather than treated as add-ons. Once children are recognised as legitimate users, it changes what digital systems must assume, protect, and enable.

Another important point to make is that Child-fit security does not aim to eliminate the role of parents or adults in protecting children online. But it rejects the assumption that parents should function as the system’s primary security mechanism or firewall. In this paradigm, parents are seen as caregivers, guides, and sources of judgement because they often understand the child’s needs, maturity, vulnerabilities, family circumstances, and social environment in ways that systems or platforms cannot. They should not be burdened to protect children or required to compensate for unsafe system designs. A child-fit security design supports the parent in protecting the child.  

\subsection{Principles of Child-Fit Security}

\subsubsection{Treating Children as Legitimate Users}
First, recognising children as legitimate users means their presence is not treated as accidental, but it is expected. Children are foreseeable users, not attackers or misuse cases that need to be prevented or protected against. This means when a system encounters a child, it will not respond by excluding them or denying them access (e.g., age gates and parental controls). While responding this way may reduce risk, it also implies that the child is not the “real” user but a special case.  A child-fit security design anticipates the presence of a child as a user who is worthy of first-order design consideration.

\subsubsection{Supporting Children’s Agency}
Second, it means recognising that children have agency~\cite{wang202312}. When children are viewed as users, they will be seen as active users, not just passive recipients of protection. This also means not viewing them as risk subject or individuals exposed to harm who should be protected from digital environments. Being seen as active users means their choices, preferences, and values are recognised as part of digital participation. This does not mean their agency is the same as adult autonomy, but it is real for their age and development~\cite{wang202312,dumaru2024s}. A child-fit system should therefore support children’s capacity to understand, choose, consent, refuse, report, recover, and participate in age-appropriate ways.

\subsubsection{Recognising and Designing for Child Diversity}
Third, it means recognising and designing for child diversity. Similar to adults, there is no single child user, every child is different. A 5 year old, 10 year old, and 15 year old are not the same, they have different needs, capabilities, and vulnerabilities. Children also differ by ability, gender, language, culture, family context, socioeconomic status and experience of risk. Recognising them as legitimate users means accounting for these differences and moving away from a generic and abstract idea of ``a child'' to a more inclusive definition of a user who is a child. A child-fit system uses these differences and acknowledges diverse interactions (or lack of) to shape security requirements. 

\subsubsection{Protecting Children's Rights}
Lastly, it means recognising that children have rights. A child as a legitimate user is entitled to the same legal protections and claims afforded to adults even more because they are children. They have the right to privacy, safety, expression, access to information, play, education, and participation. As users, these rights are not secondary to those of adults or ambitions and goals of technology platforms. While these rights should be balanced to help children develop, they cannot be ignored or silenced through system design. Consequently, a child-fit system provides an environment that recognises these rights and mechanisms that allow children to express them.

\begin{table*}[!htp]
\centering
\footnotesize
\renewcommand{\arraystretch}{1.2}
\caption{How child-fit security extends existing security, privacy, safety, and usability paradigms for children.}
\Description{This table maps five existing paradigms to their contributions, limitations in children’s digital contexts, and the additional perspective introduced by child-fit security.}
\label{tab:mapping}
\begin{tabular}{p{2.2cm}|p{4.0cm}|p{5cm}|p{5.1cm}}
\hline
\textbf{Existing Paradigm} & \textbf{What it contributes} & \textbf{Limitations when applied to children} & \textbf{What Child-fit security adds}\\
\hline
User-Centred Security & Shifts security design from purely technical systems toward the needs, practices, and limitations of users. & The ``user'' is often left underspecified and may implicitly resemble an adult characteristics such as autonomous, literate, individually responsible, and ability to understand security consequences. & Makes children explicit security subjects with age-specific capacities, dependencies, rights, relationships, and contexts of use. \\
\hline
Threat Modelling & Identifies assets, adversaries, threats, vulnerabilities, and mitigations. & Sometimes it is used to position children as weak links, misuse cases, bypassers of rules, or risks to be controlled, rather than as legitimate users to be protected.& Repositions threat modelling around children’s assets, harms, capabilities, dependencies, adversaries, and rights. \\
\hline
Privacy by Design & Embeds privacy protections into systems from the outset rather than treating privacy as an afterthought. & Children’s security needs are more than data protection, parental consent, or privacy settings alone. & Privacy is treated as one part of a broader child security requirements that also includes safety, agency, identity, relationships, participation, and wellbeing. \\
\hline
Safety by Design & Focuses on anticipating and reducing foreseeable harms in digital systems. & Improves safety of systems but can sometimes be misunderstood or used to view children primarily as vulnerable subjects to be protected, rather than as participants with agency and evolving capacities. & Balances protection with enablement and participation. It argues that children should be protected from harm while also being supported to participate, learn, explore, and seek help securely.\\
\hline
Usable Security & Seeks to make security mechanisms understandable, actionable, and less burdensome. & A mechanism that is usable for adults may still be developmentally inappropriate, confusing, frightening, or inaccessible for children. & Asks whether warnings, settings, authentication, reporting, consent flows, and recovery mechanisms are meaningful for children at different developmental stages. \\
\bottomrule
\end{tabular}
\end{table*}

\subsection{Operationalising Child-Fit Security Paradigm}
Translating child-fit security from principle into practice requires more than recognising children as a protected group, but embedding children's best interests into the everyday processes through which digital systems are designed, developed, assessed, and managed. Operationalising child-fit security means turning its principles into concrete design choices, technical safeguards, organisational practices, and evaluation criteria. This section sets out how the child-fit security paradigm can be put into practice.

\subsubsection{Childhood as a Core Design Context} 
Operationalising child-fit security begins with treating childhood as a core design context. This means treating the interest of a child as a design requirement than as a legal obligation, ethical add-on, or ethical compliance. A child-fit system should not be designed for adults first and then adjusted with child-safety features afterwards. When a system is built to prioritise child’s best interest, safety stops being an afterthought and children’s welfare is operationalised in the development of the system. This requires designers and developers to recognise that children experience digital systems differently from adults. Their cognitive, emotional, social, relational, and developmental needs should therefore shape the system’s core architecture, including its defaults, interfaces, data practices, safety mechanisms, and engagement features. In practice, this means that decisions about security, design, and engagement should not be guided only by adult-oriented metrics such as retention, monetisation, convenience, or friction reduction. These priorities must be balanced against children’s welfare. A system may be engaging or profitable, but still be inappropriate if it encourages compulsive use, exposes children to harm, relies on manipulative design, or collects more data than is necessary.

\subsubsection{Protection of Participation}
Operationalising child-fit security also means designing systems that protect children’s participation, not only systems that protect children from risk. Children’s participation in digital life should be treated as a developmental need, rather than simply as a vulnerability to be controlled or restricted. A child-fit approach recognises that children need access to digital spaces for play, learning, friendship, creativity, information, self-expression, and gradual independence. If protection is understood only as limiting access or preventing exposure to risk, it can unintentionally deny children agency and reduce the opportunities that digital technologies can provide. This means protection and participation should not be treated as competing goals. In a child-fit security paradigm, they are complementary. Systems should be designed to make children’s participation safer, more meaningful, and developmentally appropriate, rather than simply blocking, monitoring, or restricting their engagement. In practice, this requires systems to create conditions that allow children to explore and interact online with confidence. This may include age-appropriate interfaces, clear explanations, supportive safety prompts, trusted reporting mechanisms, privacy-protective defaults, graduated permissions, and safeguards against harmful contact, manipulation, or exploitation. Child-fit security therefore promotes children’s autonomy and agency while recognising their need for care and protection. The aim is not to remove children from digital life, but to design digital environments in which they can participate, develop, and exercise increasing independence safely.

\subsubsection{Protection without Surveillance.}
The existing paradigm relies on surveillance for protection; many child protective mechanisms protect children by watching them. While these mechanisms may be introduced in the name of safety, they can also reproduce the very risks that child-fit security seeks to avoid. They can undermine children’s privacy, autonomy, dignity, and trust which is essential for the protection of children. This approach can often lead to new risks such as loss of privacy, chilling effects and children exploring unsafe alternatives which may lead more risks. A child-fit security paradigm argues that protection should not come with the cost of surveillance, protection should be part of the architectural design. Safe spaces over surveillance. This means designing systems with protective defaults, data minimisation, safer interaction patterns, age-appropriate controls, limits on risky features, and clear routes for support. The aim is not to reject all forms of monitoring in every circumstance, especially where serious safeguarding risks arise. But, that surveillance should not be the primary foundation of child protection. 

\subsubsection{Developmental Appropriateness by Default} 
Child-fit security should treat age and developmental appropriateness as part of security itself, similar to encryption or access control. Protection should therefore be appropriate to children's age, maturity, and developmental needs by default, rather than added as a separate child-safety adjustment. Platforms should be designed to ``grow with the child," offering different protections, freedoms, and forms of support for children at different stages of development. Children do not belong to one category. This means security should not be blanket or uniform across all age groups, but it should be staged, measured, and developmental as children grow. For instance, for younger children, child-fit security requires stricter safeguards, such as strong privacy defaults, limited contact, restricted discoverability, and reduced exposure to persuasive or risky features. For children in middle childhood, systems should provide more guided forms of participation, including supervised choices, clear explanations, and limited agency within safe boundaries. A one-size-fits-all approach risks either being overprotective and limiting children development, or being too lenient putting children at risk.    

\subsubsection{Protection against Extractive Models and Features}
A system is not child-fit if it exploits children’s developmental vulnerabilities. Child-fit security promotes design features and algorithms that protect developmental or wellbeing risks. This is important for features designed to maximise attention, engagement, or passive consumption. Features like infinite scrolls, autoplay, streaks, push notifications and social comparison metrics may affect the development and wellbeing of children therefore should not be introduced before they are mental developed to resist or understand them. Moreover, child-fit security therefore requires platforms to assess whether such features are appropriate for children before they are introduced. Social media that may be accessed by children should prioritise child agency over passive consumption. For instance, rather pushing content to children, they should empower children to actively choose what they wand engage with. The goal here is to ensure that children’s attention, data, emotions, and relationships are not treated as resources to be extracted, but as interests to be protected.

\subsubsection{Participatory Design}
Child-fit security cannot be operationalised for children without also listening to children. It requires children’s voices to be included in the design, regulation, and evaluation of online safety measures. Children have rights and are active participants in digital life, their experiences can reveal risks, needs, and forms of support that adult designers, regulators, or policymakers may overlook. In practice, this means involving children in age-appropriate and meaningful ways when designing systems likely to affect them. Children's views should inform interface design, safety tools, reporting mechanisms, privacy settings, recommender systems, and explanations of risk. Their participation should not be treated as a symbolic after key decisions have already been made. Moreover, participatory design must itself be child-fit. Children should not be burdened with responsibility for making systems safe, their involvement should be supported, inclusive, and developmentally appropriate. The aim is to ensure that child-fit security is shaped not only by adult assumptions about children’s needs, but also by children’s lived experiences, capacities, and expectations of safety, privacy, agency, and participation.

\subsubsection{Extending Existing Threat-Modelling Techniques}
Operationalising child-fit security means extending existing security and privacy techniques so that they better account for children’s needs, rights, and developmental capacities. It does not require replacing established methods such as STRIDE or LINDDUN. Child-fit builds on them by ensuring that child-specific risks and protections are considered as part of ordinary security and privacy engineering practice. Frameworks such as STRIDE and LINDDUN already provide useful ways to identify security and privacy threats. A child-fit approach strengthens these techniques by asking how each threat might affect children differently, and what additional safeguards may be needed when children are likely users of a system. For example, information disclosure should include not only unauthorised access to data, but also risks involving a child’s identity, location, relationships, vulnerabilities, or private expression. Linkability should include the possibility that children may be profiled across different contexts over time. Unawareness should include whether children can realistically understand how their data is collected, how recommendations work, or who may interact with them. The aim is not to create an entirely separate security process, but to make child-fit considerations part of the standard process. Child-fit security therefore operationalises children’s best interests by extending existing threat-modelling techniques to ensure that children’s safety, dignity, privacy, agency, and development are considered from the outset.

\subsubsection{Parent-in-the-Loop}
Child-fit security recognises the role that parents and caregivers play in children’s digital lives. They often hold contextual knowledge that platforms cannot easily infer, including a child’s maturity, vulnerabilities, relational context, and support needs~\cite{akter2022parental,dumaru2024s}. A parent-in-the-loop approach here means that systems should create meaningful points at which parents and caregivers can support children’s digital participation. This may include helping children understand risks, adjust settings, or report problems~\cite{zhao2019make,wang202312}. Such involvement helps translate child safety protections into support that fits the individual child.  However, parent-in-the-loop should not become parent-as-surveillance. The aim is not to make parents constantly monitor children’s activity, but to involve them where their judgement or care is needed. Child-fit security therefore requires parental involvement to be timely, proportionate, privacy-respecting, and developmentally sensitive. Overall, parent-in-the-loop operationalises child-fit security by connecting system-level safeguards with the lived realities of children’s lives. It ensures that parents and caregivers can support fit without being made responsible for compensating for unsafe design.

\begin{figure}[!htp]
    \centering
    \includegraphics[width=\linewidth]{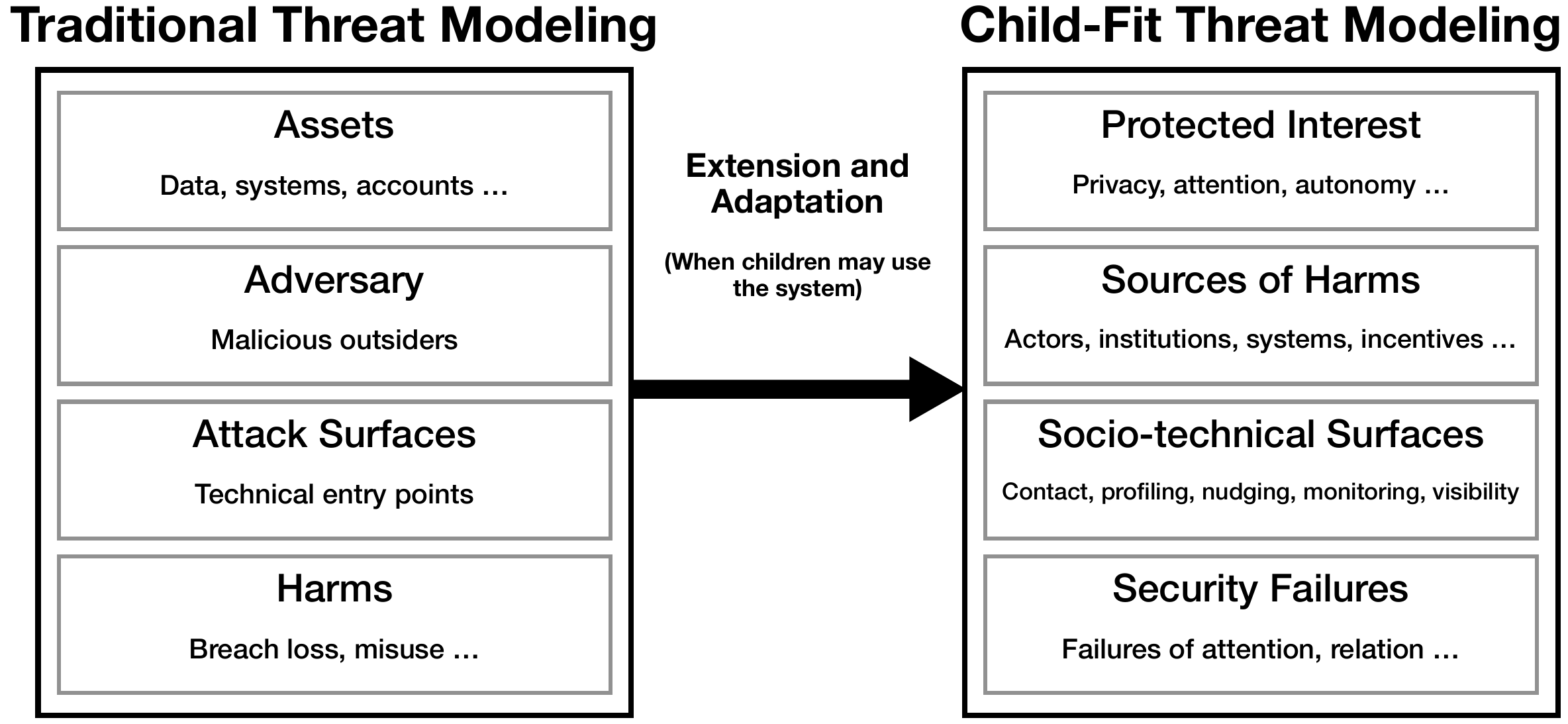}
    \caption{When children are involved, we propose that designers rethink how they conduct their threat modelling.}
    \Description{Threat modelling should extend and cover children.}
    \label{fig:threatmodeling}
\end{figure}

\subsection{Threat Modelling For Child-Fit Technology}
A child-fit approach to technology requires an extension of existing threat-modelling techniques. Many threat-modelling techniques already ask valuable security questions such as what is being protected, what can go wrong, who might cause harm, and how risks might be mitigated. These categories remain useful when protecting children. However, extending them to children requires rethinking of what counts as an asset worth protecting. The aim of this section is not to just add children to existing threat models, but to show how threat modelling itself must be adapted when children may use digital systems. The examples we use here are not exhaustive list of potential assets, harms, and threats. Figure~\ref{fig:threatmodeling} visualises how we conceptualise the extension of existing threat modelling techniques.  

\subsubsection{Assets.} This is because children are not only users, account holders, or data subjects. They are persons whose privacy, autonomy, attention, wellbeing, and relationships can be shaped by digital systems. In child-fit threat modelling, these are assets that can be harmed even when the system is not hacked or unauthorised access.

\subsubsection{Adversary.} The idea of an adversary should also move beyond intentional actors, such as groomers or abusive peers, but ignorant or opportunistic actors such as peers who may unintentionally amplify shame or parents who misuse monitoring tools. In addition, harm may also arise from structural sources such as recommender systems, advertising systems, and dark patterns.

\subsubsection{Attack Surfaces.} The concept of the attack surface also needs to be reinterpreted. In child-fit threat modelling, an attack surface is not only a technical point of entry into a system but could be any interface, feature, or process through which a child can be contacted, monitored or made visible to others. Obvious surfaces may include public profiles, livestreaming, and tagged content, while less obvious surfaces can include influencer content, age assurance systems, and parental dashboards. This is important because some features are often presented as protective mechanisms but can also generate child-specific risks. For instance, parental apps may support safety in some contexts, but they can also enable coercive monitoring or prevent children from seeking help. 

\subsubsection{Harms.} Child-fit thinking also reframes how outcomes are understood. Harms are not just understood as system outputs but also as security failures. Framing harms as security failures helps designers, regulators, and researchers see that child safety is not external to security. The main idea here is to change the way threat modelling understands what is valuable, where risk arises, who or what can produce harm, and what counts as a system failure. 

\subsubsection{Misuse Cases.} Misuse cases are also critical to this paradigm. Their inclusion can show how ordinary system features may combine to produce harm. For instance, an AI companion may encourage repeated disclosure or dependency because it is optimised for engagement.

\section{Case Illustration}

In this section, we illustrate how the four principles of \emph{Child-Fit Security} could be applied in practice. Rather than developing a complete platform design, we focus on showing how child protection and agency may be supported across childhood when children are treated as legitimate users (see Table~\ref{tab:case}).

\begin{framed}
    \textit{\textbf{Scenario:} We consider a hypothetical social media platform used by three children at different stages of development: a 6-year-old, a 10-year-old, and a 16-year-old. The platform includes profile creation, group participation, feed-based content, sharing, direct messaging, and search.}
\end{framed}

\begin{table*}[ht]
\centering
\footnotesize
\renewcommand{\arraystretch}{1.8}
\caption{Illustration of \emph{Child-Fit Security} across three stages of childhood.}
\label{tab:case}
\begin{tabular}{p{2.0cm}|p{4.5cm}|p{4.5cm}|p{4.5cm}}
\toprule
    \textbf{Principle} & \textbf{6-year-old (Scaffolded Protection)} & \textbf{10-year-old (Guided Protection)} & \textbf{16-year-old (Supported Protection)} \\
\hline
    \textbf{Treating children as legitimate users} &
    
    \textbf{Design:} Recognised as a user on play and discovery. Designed for them from the outset.
    
    \textbf{Operationalisation:} 
    \begin{itemize} [leftmargin=*, noitemsep, topsep=1pt]
    \item Adult-supported account setup
    \item Adults acting as gatekeepers
    \end{itemize}
&
    \textbf{Design:} Recognised as a user with growing social and learning needs.
    
    \textbf{Operationalisation:} 
    \begin{itemize} [leftmargin=*, noitemsep, topsep=1pt]
    \item Protective settings remain on
    \item Access to approved groups/activities
    \end{itemize}
&
    \textbf{Design:} Recognised as a legitimate user with stronger privacy and participation needs.

    \textbf{Operationalisation:}
    \begin{itemize}[leftmargin=*, noitemsep, topsep=1pt]
    \item Access to wider community groups
    \item Adults acting as partner
    \end{itemize} 
\\
\hline
    \textbf{Supporting children’s agency} &
    \textbf{Design:} Simple choices within safe boundaries. Adults remain clearly in the loop.

    \textbf{Operationalisation:}
    \begin{itemize}[leftmargin=*, noitemsep, topsep=1pt]
    \item Visual interface for controls
    \item Voice-messaging function replacing typing
    \end{itemize}
&
    \textbf{Design:} Adults help with safety boundaries but do not always make decisions.

    \textbf{Operationalisation:}
    \begin{itemize}[leftmargin=*, noitemsep, topsep=1pt]
    \item Child suggests preferences (e.g., when to use the platform)
    \item Adults confirm broad boundaries
    \end{itemize}  
&
    \textbf{Design:} Transparency and direct control. Adults provide support not constant oversight.

    \textbf{Operationalisation:}
    \begin{itemize}[leftmargin=*, noitemsep, topsep=1pt]
    \item User can view and adjust settings
    \item Adults informed via a ``Safety Report'' rather than direct surveillance of DMs
    \end{itemize}
\\
\hline
    \textbf{Recognising and designing for child diversity} &
    \textbf{Design:} Simple and child-friendly interface.

    \textbf{Operationalisation:}
    \begin{itemize}[leftmargin=*, noitemsep, topsep=1pt]
    \item Larger buttons for small fingers that might lack fine precision
    \item No ``likes'' that incentivise addictive loops
    \end{itemize}
&
    \textbf{Design:} Flexible child protection due to different literacy and maturity levels in this age.

    \textbf{Operationalisation:}
    \begin{itemize}[leftmargin=*, noitemsep, topsep=1pt]
    \item Adjustable explanations
    \item A break function to interrupt infinite scroll
    \end{itemize}  
&
    \textbf{Design:} Adaptable controls and content filtering for different capacities and vulnerabilities.
    
    \textbf{Operationalisation:}
    \begin{itemize}[leftmargin=*, noitemsep, topsep=1pt]
    \item Flexible support options
    \item User can tweak their interesting content
    \end{itemize}
\\
\hline
    \textbf{Protecting children's rights} &
    \textbf{Design:} Rights preserved within highly protected boundaries.
        
    \textbf{Operationalisation:}
    \begin{itemize}[leftmargin=*, noitemsep, topsep=1pt]
    \item Security and privacy on by default
    \item Simple block/stop functions
    \end{itemize}
&
    \textbf{Design:} Rights preserved within participation. Informed when controls or rules change.

    \textbf{Operationalisation:}
    \begin{itemize}[leftmargin=*, noitemsep, topsep=1pt]
    \item Notifications trigger a conversation rather than just a block
    \item Safer communication
    \end{itemize}
&
    \textbf{Design:} Rights preserved within dignity, trust, and autonomy.

    \textbf{Operationalisation:}
    \begin{itemize}[leftmargin=*, noitemsep, topsep=1pt]
    \item Ability to challenge filters (e.g., ``requesting an override'' from a parent)
    \item ``forget'' content from younger years
    \end{itemize}
\\
\bottomrule
\end{tabular}
\end{table*}

\subsection{Motivating scenario}

\begin{framed}

Alice and Dave are both children growing up in different regulatory environments.
    
Alice lives in a country that does not simply ban children from social media. Apps and websites that children are likely to use are required to be designed with children in mind.
Ages 3–5: Alice watches videos on her parents' devices through a child-friendly platform. The content is carefully selected, recommendations are limited, and her parents are closely involved. They help her understand what she is seeing.

Ages 6–10: Alice starts using more apps. She picks some videos herself, plays games, and begins making small choices but always within systems built for her age group, and with her parents' permission and support. The goal is not unrestricted access, but supported exploration.

Ages 10–13: Alice wants to chat with friends and join small online communities. The platform she uses lets her connect with classmates and friends from school or local clubs, but strangers cannot find her profile or message her. She can add or remove friends, report problems, and ask for help if something feels wrong. Her parents do not need to monitor her every move they are there when she needs them.

By 16: Alice joins a popular social media platform used by her peers. She is not risk-proof, but she had exposure to  managing her contacts, spotting uncomfortable situations, using privacy settings, and knowing when to talk to a trusted adult.

Dave grows up in a country where children are banned from social media until they turn 16. Before that, his online life is mostly games and apps his parents have approved. When he turns 16, he joins the same platform as his friends, but this is his first time using large-scale social media. Unlike Alice, he has not had the chance to build up habits, coping skills, or the confidence to ask for help when something goes wrong.

\end{framed}
\textbf{\emph{The question is not which child is completely safe. Neither is. The question is which child has been better prepared for meaningful and safer participation. Alice’s pathway reflects a child-fit model: protection is built into the environment while participation gradually expands. Dave’s pathway reflects a containment model: risk is delayed, but not necessarily reduced, and the child may enter adult-scale digital spaces with little preparation.}}

\section{Research Agenda}
Child-fit security introduces a design paradigm in which technologies likely to be used by children treat children's wellbeing, development, privacy, safety, agency, and rights as core security requirements. However, operationalising this paradigm introduces several conceptual, technical, empirical, and governance challenges. In this section, we outline a research agenda for developing the empirical basis, design methods, technical mechanisms, and accountability structures needed to realise child-fit security in practice. 

\subsection{Inclusion of Children in the Design of Security and Protection}
Most child-protection mechanisms are usually designed by adults, for children, and then imposed on them through restrictions and surveillance~\cite{hartikainen2019childrensdesign,Kinnula2021manifesto}. Children are treated as passive recipients of protection, while their voices are missing. For instance, they are rarely involved in defining and shaping how protection should be designed and experienced in ways that are appropriate to their daily lives~\cite{Liu2025codesign,Badillo2019stranger}. This creates a significant gap: adults may misjudge what harms matter most and what protections feel helpful while children often lack opportunities to contribute their own views, issues, interests, and opinions to online security design~\cite{williams2023youth,hartikainen2019childrensdesign,zhao2019make,Badillo2019stranger}.

\textbf{\textit{Research Agenda:}} Future research should investigate how children can be meaningfully included in the design of security and protection while avoiding tokenistic inclusion. This could include understanding how their views and opinions can be used to shape the design of practical security mechanisms. Advancing this work would help move child protection away from adult-imposed security control and toward systems that are more legitimate, usable, and responsive to children’s lived experiences.

\subsection{Designing for Different Ages and Developmental Stages}
Security design often treats children as a single group, for example, regulations may draw an age borderline such as 13, with parental controls or consent required below that age and fewer protections for children over 13 using online services like social media, gaming, or app sign-ups~\cite{coppa,crc_gc25}. However, children should not be treated as a single group. A 5-year-old and a 10-year-old have different capacities, developmental needs, and ways of accessing technology, yet many current protections remain same~\cite{zhao2019silly}. Consequently, a one-size-fits-all approach has limitations~\cite{crc_gc25,ico_code,williams2023youth}. A major challenge is therefore to translate developmental appropriateness into design and security requirements.

\textbf{\textit{Research Agenda:}} Future research should investigate how developmental stages can inform security requirements, defaults, permissions, explanations, and reporting mechanisms. This should be accompanied by age-sensitive designs that support gradual autonomy to avoid having blanket bans and exclusions. We acknowledge that this would come with debates around what is appropriate, but research should explore how children, parents, educators, and child-development experts define ``appropriate'' across different contexts, cultures, and risk environments. Advancing this work would help move child protection away from one-size-fits-all approaches and toward systems that better match children's evolving capacities and needs.

\subsection{Threat Modelling for Child-Fit Systems}
Children are often absent from designers' threat models because they are not fully recognised as legitimate users of digital systems~\cite{williams2023youth,kumar2023understanding}. Even when children are considered, existing threat models are narrow as they mainly focus on data risks, account compromise, access control, and privacy management, while paying less attention to children's interactions, developmental needs, and lived experiences~\cite{zhao2019silly,wang2023datafication}. As a result, we realise that important harms to wellbeing, agency, and trust are easily overlooked, even though these harms may be central to children’s digital security~\cite{wang2023datafication,crc_gc25}. We have shown that threat modelling through child-fit security `thinking' threats may arise from unexpected sources. This can complicates the boundary between adversarial threat, harmful design, and ordinary system behaviour.

\textbf{\textit{Research Agenda:}} Researchers should develop threat-modelling methods specifically for systems likely to be used by children. These methods should identify adversaries, abuse cases, design-induced harms, and institutional incentives that may undermine children's wellbeing. Future work should also examine how to model non-traditional adversaries, including recommender systems optimised for engagement, advertising infrastructures and caregivers who misuse monitoring tools. Moreover, adolescence should be included in threat modelling exercises to understand how they understand harm and threats. Prior work~\cite{theofanos2021passwords,paudel2024leveraging,zhang2026player} has shown that children can contribute to their protections.

\subsection{Protection through System Design, not Restriction and Surveillance}
Many current protections for children rely on surveillance, restriction, exclusion, or denial of access~\cite{jarvie2024age,ofcom2025agechecks}. However, these responses often fail to address the design of the systems themselves~\cite{jarvie2024age,wang2023datafication}. Children may circumvent bans through VPNs, false accounts, borrowed devices, or other workarounds~\cite{ofcom_families,yao2025age}. This creates a weak form of protection: children may still access technologies but without age-appropriate safeguarding mechanisms, transparent protections, or accountable system design. Moreover, if children are excluded from their systems, platforms or service providers may have less incentive to improve child security and safety~\cite{ofcom_age,ofcom2025agechecks,jarvie2024age}.

\textbf{\textit{Research Agenda:}} Future research should investigate how protections can be embedded into system architecture and design, rather than through surveillance, bans, exclusion, and after-the-fact control. Exploring this work could also explore alternatives to intrusive or exclusionary controls, including privacy-preserving approaches to age assurance and safety features that protect children without undermining trust, autonomy, or participation.

\subsection{Implementation and Developer Support}
Security is about tools and processes. Child-fit security will fail if it remains only a design principle. Developers need concrete tools, patterns, APIs, documentation, checklists, test suites, and engineering processes. Ekambaranathan et al.~\cite{ekambaranathan2021money} found that developers of children’s apps face barriers to better privacy, including monetisation pressures, third-party libraries, and lack of guidance. Reyes et al.~\cite{reyes2018won} also analysed 5,855 children’s apps and found widespread potential COPPA compliance issues, especially involving third-party SDKs.

\textbf{\textit{Research Agenda:}} Future work should develop engineering support for child-fit security across the software development lifecycle. This includes child-fit threat-modelling templates, misuse-case libraries, child-centred privacy labels, SDK risk analysis tools, and child-fit design review processes. Also, research should examine how child-fit requirements can be embedded into agile development, risk assessment, app-store review, and platform governance without becoming a box-ticking exercise.

\section{Conclusion}
Child online protection has long operated through a logic of control and restriction. This framing is understandable, but it is also limiting. Containment has become the default response, while many systems continue to be designed without recognising children as legitimate participants. Controls, restrictions, and bans have become the norm. To address this, we proposed a new way of conceptualising child protection. Child-fit security begins from a different premise that if children are likely to use a system, that system should be accountable for the kind of childhood it makes possible. This requires designing for protection and participation together, not as competing priorities, but as a single requirement. Children should not have to be excluded from digital life to be safe, nor exposed to adult-scale risks to participate. This reframing has consequences for how security itself is understood. A narrow conception of child protection fails to account for childhood, or for the conditions under which children learn, play, communicate, and develop. This is not the end of security, but perhaps the end of a narrow security that treats children as outsiders, risks, or edge cases.


\bibliographystyle{ACM-Reference-Format}
\bibliography{references}

@online{pewTeensInternetDeviceAccess2025,
  author       = {{Pew Research Center}},
  title        = {Teens and Internet, Device Access Fact Sheet},
  date         = {2025-07-10},
  url          = {https://www.pewresearch.org/internet/fact-sheet/teens-and-internet-device-access-fact-sheet/},
  urldate      = {2026-05-13},
  organization = {Pew Research Center}
}

@report{ofcomChildrenMediaUseAttitudes2025,
  author       = {{Ofcom}},
  title        = {Children and Parents: Media Use and Attitudes Report},
  institution  = {Ofcom},
  date         = {2025-05-07},
  url          = {https://www.ofcom.org.uk/siteassets/resources/documents/research-and-data/media-literacy-research/children/childrens-media-use-and-attitudes-report-2025/childrens-media-literacy-report-2025.pdf?v=396621},
  urldate      = {2026-05-13}
}

@report{ofcomChildrenPassiveOnlineMeasurement2025,
  author       = {{Ofcom}},
  title        = {Children's Passive Online Measurement},
  institution  = {Ofcom},
  date         = {2025-06-27},
  url          = {https://www.ofcom.org.uk/siteassets/resources/documents/online-safety/research-statistics-and-data/protecting-children/ofcom-childrens-passive-online-measurement.pdf?v=408844},
  urldate      = {2026-05-13}
}

@report{mannCommonSenseCensus2025,
  author       = {Mann, Supreet and Calvin, Angela and Lenhart, Amanda and Robb, Michael B.},
  title        = {The Common Sense Census: Media Use by Kids Zero to Eight, 2025},
  institution  = {Common Sense Media},
  location     = {San Francisco, CA},
  year         = {2025},
  url          = {https://www.commonsensemedia.org/sites/default/files/research/report/2025-common-sense-census-web-2.pdf},
  urldate      = {2026-05-13}
}

@online{bbcAustraliaSocialMediaBan2025,
  author       = {{BBC News}},
  title        = {Australia's Social Media Ban for Kids Under 16 -- How Will It Work?},
  date         = {2025-11-21},
  url          = {https://www.bbc.com/news/articles/cwyp9d3ddqyo},
  urldate      = {2026-05-13},
  organization = {BBC News}
}

@article{marwick2008catch,
  author  = {Marwick, Alice E.},
  title   = {To Catch a Predator? The MySpace Moral Panic},
  journal = {First Monday},
  volume  = {13},
  number  = {6},
  year    = {2008},
  date    = {2008-05-19},
  doi     = {10.5210/fm.v13i6.2152},
  url     = {https://firstmonday.org/ojs/index.php/fm/article/view/2152},
  urldate = {2026-05-09}
}

@misc{house2006dopa,
  author       = {{U.S. Congress, House Committee on Energy and Commerce, Subcommittee on Telecommunications and the Internet}},
  title        = {{H.R. 5319, The Deleting Online Predators Act of 2006}},
  howpublished = {Hearing before the Subcommittee on Telecommunications and the Internet, Committee on Energy and Commerce, House of Representatives, 109th Congress, 2nd Session},
  date         = {2006-07-11},
  year         = {2006},
  number       = {Serial No. 109-121},
  publisher    = {U.S. Government Printing Office},
  address      = {Washington, DC},
  url          = {https://www.govinfo.gov/content/pkg/CHRG-109hhrg30410/html/CHRG-109hhrg30410.htm},
  urldate      = {2026-05-09}
}

@misc{ukparliament2021draftonlinesafetybill,
  author       = {{UK Parliament, Joint Committee on the Draft Online Safety Bill}},
  title        = {Corrected Oral Evidence: Consideration of Government's Draft Online Safety Bill},
  howpublished = {Oral evidence, Evidence Session No. 11, Questions 154--192},
  date         = {2021-10-25},
  year         = {2021},
  note         = {Witness: Frances Haugen, former Facebook employee},
  url          = {https://committees.parliament.uk/oralevidence/2884/html/},
  urldate      = {2026-05-09}
}

@misc{zeigler2023cda230,
  author       = {Zeigler, Sara L.},
  title        = {Communications Decency Act and Section 230 (1996)},
  howpublished = {The First Amendment Encyclopedia},
  date         = {2023-05-23},
  year         = {2023},
  note         = {Last updated July 5, 2024},
  url          = {https://firstamendment.mtsu.edu/article/communications-decency-act-and-section-230/},
  urldate      = {2026-05-09}
}

@incollection{nasem2024introduction,
  author    = {{National Academies of Sciences, Engineering, and Medicine}},
  title     = {Introduction},
  booktitle = {Social Media and Adolescent Health},
  editor    = {Wojtowicz, Anne and Buckley, Gillian J. and Galea, Sandro},
  publisher = {National Academies Press},
  address   = {Washington, DC},
  year      = {2024},
  date      = {2024-03-25},
  url       = {https://www.ncbi.nlm.nih.gov/books/NBK603432/},
  urldate   = {2026-05-09}
}

@article{BARRY20171,
title = {Adolescent social media use and mental health from adolescent and parent perspectives},
journal = {Journal of Adolescence},
volume = {61},
pages = {1-11},
year = {2017},
issn = {0140-1971},
doi = {https://doi.org/10.1016/j.adolescence.2017.08.005},
url = {https://www.sciencedirect.com/science/article/pii/S0140197117301318},
author = {Christopher T. Barry and Chloe L. Sidoti and Shanelle M. Briggs and Shari R. Reiter and Rebecca A. Lindsey}
}

@article{abelson2024bugs,
  title={Bugs in our pockets: the risks of client-side scanning},
  author={Abelson, Harold and Anderson, Ross and Bellovin, Steven M and Benaloh, Josh and Blaze, Matt and Callas, Jon and Diffie, Whitfield and Landau, Susan and Neumann, Peter G and Rivest, Ronald L and others},
  journal={Journal of Cybersecurity},
  volume={10},
  number={1},
  pages={tyad020},
  year={2024},
  publisher={Oxford University Press}
}

@inproceedings{struppek2022learning,
  title={Learning to break deep perceptual hashing: The use case neuralhash},
  author={Struppek, Lukas and Hintersdorf, Dominik and Neider, Daniel and Kersting, Kristian},
  booktitle={Proceedings of the 2022 ACM Conference on Fairness, Accountability, and Transparency},
  pages={58--69},
  year={2022}
}

@misc{esafety_social_media_age_restrictions_2026,
  author       = {{eSafety Commissioner}},
  title        = {{Social Media Age Restrictions}},
  year         = {2026},
  url          = {https://www.esafety.gov.au/about-us/industry-regulation/social-media-age-restrictions},
  note         = {Accessed 9 May 2026}
}

@misc{esafety_age_restricted_platforms_2026,
  author       = {{eSafety Commissioner}},
  title        = {{Which Social Media Platforms Are Age-Restricted?}},
  year         = {2026},
  month        = mar,
  day          = {30},
  url          = {https://www.esafety.gov.au/about-us/industry-regulation/social-media-age-restrictions/which-platforms-are-age-restricted},
  note         = {Accessed 9 May 2026}
}

@misc{oaic_social_media_minimum_age_2025,
  author       = {{Office of the Australian Information Commissioner}},
  title        = {{Social Media Minimum Age}},
  year         = {2025},
  month        = oct,
  day          = {23},
  url          = {https://www.oaic.gov.au/privacy/your-privacy-rights/social-media-minimum-age},
  note         = {Accessed 9 May 2026}
}

@online{demarest_eca_digital_2026,
  author       = {{Demarest Advogados}},
  title        = {{Digital Statute for Children and Adolescents -- Law No. 15,211/2025}},
  year         = {2026},
  month        = mar,
  day          = {18},
  url          = {https://www.demarest.com.br/en/digital-statute-for-children-and-adolescents-law-no-15211-2025/},
  note         = {Accessed 9 May 2026}
}

@online{baker_mckenzie_digital_eca_2026,
  author       = {{Baker McKenzie}},
  title        = {{Brazil Regulates the Children and Adolescents Online Safety Act}},
  year         = {2026},
  month        = mar,
  day          = {19},
  url          = {https://www.bakermckenzie.com/en/insight/publications/2026/03/brazil-regulates-the-children-and-adolescents-online-safety-act},
  note         = {Accessed 9 May 2026}
}

@misc{freedom_house_china_fotn_2024,
  author       = {{Freedom House}},
  title        = {{China: Freedom on the Net 2024 Country Report}},
  year         = {2024},
  url          = {https://freedomhouse.org/country/china/freedom-net/2024},
  note         = {Accessed 9 May 2026}
}

@misc{china_state_council_online_gaming_2021,
  author       = {{The State Council of the People's Republic of China}},
  title        = {{Stricter Limits on Minors' Online Gaming}},
  year         = {2021},
  month        = aug,
  day          = {31},
  url          = {https://english.www.gov.cn/statecouncil/ministries/202108/31/content_WS612da2a2c6d0df57f98df6bd.html},
  note         = {Accessed 9 May 2026}
}

@misc{dcms_homeoffice_2021_safetytech,
  author       = {{Department for Digital, Culture, Media \& Sport} and {Home Office}},
  title        = {Government funds new tech in the fight against online child abuse},
  year         = {2021},
  month        = nov,
  day          = {17},
  howpublished = {\url{https://www.gov.uk/government/news/government-funds-new-tech-in-the-fight-against-online-child-abuse}},
  note         = {Accessed 10 May 2026}
}

@report{internetsociety_2023_clientsidescanning,
  author       = {{Internet Society}},
  title        = {Client-Side Scanning: What It Is and Why It Threatens Trustworthy, Private Communication},
  year         = {2023},
  month        = dec,
  day          = {13},
  institution  = {Internet Society},
  url          = {https://www.internetsociety.org/resources/doc/2023/client-side-scanning/},
  note         = {Accessed 10 May 2026}
}

@misc{reuters_2025_appleadpuk,
  author       = {{Reuters}},
  title        = {Apple pulls data protection feature in UK amid government demands},
  year         = {2025},
  month        = feb,
  day          = {21},
  howpublished = {\url{https://www.reuters.com/technology/apple-removing-end-to-end-cloud-encryption-feature-uk-bloomberg-news-reports-2025-02-21/}},
  note         = {Accessed 10 May 2026}
}

@report{unicef_2020_encryptionchildren,
  author       = {{UNICEF Innocenti}},
  title        = {Encryption, Privacy and Children’s Right to Protection from Harm},
  year         = {2020},
  institution  = {UNICEF Innocenti},
  url          = {https://www.unicef.org/innocenti/media/3446/file/UNICEF-Encryption-Privacy-Right-Protection-From-Harm-2020.pdf}
}

@misc{apple_2025_adpuk,
  author       = {{Apple}},
  title        = {Apple can no longer offer Advanced Data Protection in the United Kingdom to new users},
  year         = {2025},
  month        = sep,
  day          = {23},
  howpublished = {\url{https://support.apple.com/en-gb/122234}},
  note         = {Accessed 10 May 2026}
}

@report{unicef2025childhooddigitalworld,
  author      = {{UNICEF Innocenti}},
  title       = {Childhood in a Digital World},
  institution = {UNICEF Innocenti -- Global Office of Research and Foresight},
  year        = {2025},
  month       = jun,
  day         = {12},
  url         = {https://www.unicef.org/innocenti/reports/childhood-digital-world}
}

@misc{newamerica2023antilgbtqfilters,
  author       = {{New America}},
  title        = {How Anti-LGBTQ Web Filters Stand Between LGBTQ Youth and the Online Resources They Need},
  year         = {2023},
  howpublished = {\url{https://www.newamerica.org/insights/how-anti-lgbtq-web-filters-stand-between-lgbtq-youth-and-the-online-resources-they-need/}}
}

@misc{ofcom2025agechecks,
  author       = {{Ofcom}},
  title        = {Age Checks to Protect Children Online},
  year         = {2025},
  month        = jan,
  day          = {16},
  howpublished = {\url{https://www.ofcom.org.uk/online-safety/protecting-children/age-checks-to-protect-children-online}}
}

@article{helsper2021digitalinequalities,
  author  = {Helsper, Ellen J.},
  title   = {The Digital Disconnect: The Social Causes and Consequences of Digital Inequalities},
  journal = {SAGE Publications},
  year    = {2021},
  url     = {https://uk.sagepub.com/en-gb/eur/the-digital-disconnect/book270303}
}

@misc{unicef2025bestinterestsdigital,
  author      = {{UNICEF Innocenti}},
  title       = {Best Interests of the Child in Relation to the Digital Environment},
  institution = {UNICEF Innocenti -- Global Office of Research and Foresight},
  year        = {2025},
  url         = {https://www.unicef.org/innocenti/reports/best-interests-child-relation-digital-environment}
}

@article{hartikainen2019childrensdesign,
  author  = {Hartikainen, Heidi and Iivari, Netta and Kinnula, Marianne},
  title   = {Children's Design Recommendations for Online Safety Education},
  journal = {International Journal of Child-Computer Interaction},
  volume  = {22},
  year    = {2019},
  pages   = {100146},
  doi     = {10.1016/j.ijcci.2019.100146},
  url     = {https://www.sciencedirect.com/science/article/abs/pii/S2212868917300764}
}

@article{shaffique2025behaviouralprofiling,
  author  = {Shaffique, Muhammad R. and van der Hof, Simone},
  title   = {Behavioural Profiling for Age Assurance: Do the Ends Justify the Means?},
  journal = {International Data Privacy Law},
  year    = {2025},
  doi     = {10.1093/idpl/ipaf012},
  url     = {https://doi.org/10.1093/idpl/ipaf012}
}

@article{jarvie2024onlineageverification,
  author  = {Jarvie, Chelsea and Renaud, Karen},
  title   = {Online Age Verification: Government Legislation, Supplier Responsibilization, and Public Perceptions},
  journal = {Children},
  volume  = {11},
  number  = {9},
  pages   = {1068},
  year    = {2024},
  doi     = {10.3390/children11091068},
  url     = {https://doi.org/10.3390/children11091068}
}

@article{notley2009young,
  title={Young people, online networks, and social inclusion},
  author={Notley, Tanya},
  journal={Journal of computer-mediated communication},
  volume={14},
  number={4},
  pages={1208--1227},
  year={2009},
  publisher={Oxford University Press Oxford, UK}
}

@article{Kinnula2021manifesto,
    title     = {Manifesto for children’s genuine participation in digital technology design and making},
    journal   = {International Journal of Child-Computer Interaction},
    volume    = {28},
    pages     = {100244},
    year      = {2021},
    issn      = {2212-8689},
    doi       = {https://doi.org/10.1016/j.ijcci.2020.100244},
    author    = {Marianne Kinnula and Netta Iivari}
}

@inproceedings{Liu2025codesign,
    author      = {Liu, Lanjing and Wang, Xiaozheng and Hasan, Shaddi and Yao, Yaxing},
    title       = {Co-Design Privacy Notice and Controls with Children},
    year        = {2025},
    isbn        = {9798400713958},
    publisher   = {Association for Computing Machinery},
    address     = {New York, NY, USA},
    url         = {https://doi.org/10.1145/3706599.3719886},
    doi         = {10.1145/3706599.3719886},
    booktitle   = {Proceedings of the Extended Abstracts of the CHI Conference on Human Factors in Computing Systems},
    articleno   = {137},
    numpages    = {7},
    series      = {CHI EA '25}
}

@inproceedings{Badillo2019stranger,
    author      = {Badillo-Urquiola, Karla and Smriti, Diva and McNally, Brenna and Golub, Evan and Bonsignore, Elizabeth and Wisniewski, Pamela J.},
    title       = {Stranger Danger! Social Media App Features Co-designed with Children to Keep Them Safe Online},
    year        = {2019},
    isbn        = {9781450366908},
    publisher   = {Association for Computing Machinery},
    address     = {New York, NY, USA},
    url         = {https://doi.org/10.1145/3311927.3323133},
    doi         = {10.1145/3311927.3323133},
    booktitle   = {Proceedings of the 18th ACM International Conference on Interaction Design and Children},
    pages       = {394–406},
    numpages    = {13},
    location    = {Boise, ID, USA},
    series      = {IDC '19}
}

@misc{coppa,
    author      = {{Federal Trade Commission}},
    title       = {Children's Online Privacy Protection Rule ("COPPA")},
    year        = {n.d.},
    howpublished= {\url{https://www.ftc.gov/legal-library/browse/rules/childrens-online-privacy-protection-rule-coppa}}
}

@misc{crc_gc25,
    author      = {{UN Committee on the Rights of the Child}},
    title       = {General Comment No. 25 (2021) on Children's Rights in Relation to the Digital Environment},
    year        = {2021},
    howpublished= {\url{https://www.ohchr.org/en/documents/general-comments-and-recommendations/general-comment-no-25-2021-childrens-rights-relation}}
}

@inproceedings{williams2023youth,
    author      = {Olivia Williams and Yee-Yin Choong and Kerrianne Buchanan},
    title       = {Youth understandings of online privacy and security: A dyadic study of children and their parents},
    booktitle   = {Nineteenth Symposium on Usable Privacy and Security (SOUPS 2023)},
    year        = {2023},
    isbn        = {978-1-939133-36-6},
    address     = {Anaheim, CA},
    pages       = {399--416},
    url         = {https://www.usenix.org/conference/soups2023/presentation/williams},
    publisher   = {USENIX Association}
}

@inproceedings{zhao2019silly,
    author      = {Zhao, Jun and Wang, Ge and Dally, Carys and Slovak, Petr and Edbrooke-Childs, Julian and Van Kleek, Max and Shadbolt, Nigel},
    title       = {`I make up a silly name': Understanding Children's Perception of Privacy Risks Online},
    year        = {2019},
    isbn        = {9781450359702},
    publisher   = {Association for Computing Machinery},
    address     = {New York, NY, USA},
    url         = {https://doi.org/10.1145/3290605.3300336},
    doi         = {10.1145/3290605.3300336},
    booktitle   = {Proceedings of the 2019 CHI Conference on Human Factors in Computing Systems},
    pages       = {1–13},
    location    = {Glasgow, Scotland Uk},
    series      = {CHI '19}
}

@misc{ico_code,
    author      = {{Information Commissioner's Office}},
    title       = {Age Appropriate Design: A Code of Practice for Online Services},
    year        = {n.d.},
    howpublished= {\url{https://ico.org.uk/for-organisations/uk-gdpr-guidance-and-resources/childrens-information/childrens-code-guidance-and-resources/age-appropriate-design-a-code-of-practice-for-online-services/}}
}

@inproceedings{wang2023datafication,
    author      = {Wang, Ge and Zhao, Jun and Van Kleek, Max and Shadbolt, Nigel},
    title       = {‘Treat me as your friend, not a number in your database’: Co-designing with Children to Cope with Datafication Online},
    year        = {2023},
    isbn        = {9781450394215},
    publisher   = {Association for Computing Machinery},
    address     = {New York, NY, USA},
    url         = {https://doi.org/10.1145/3544548.3580933},
    doi         = {10.1145/3544548.3580933},
    booktitle   = {Proceedings of the 2023 CHI Conference on Human Factors in Computing Systems},
    articleno   = {95},
    location    = {Hamburg, Germany},
    series      = {CHI '23}
}

@inproceedings{kumar2023understanding,
    author      = {Kumar, Priya C. and O'Connell, Fiona and Li, Lucy and Byrne, Virginia L. and Chetty, Marshini and Clegg, Tamara L. and Vitak, Jessica},
    title       = {Understanding Research Related to Designing for Children's Privacy and Security: A Document Analysis},
    year        = {2023},
    isbn        = {9798400701313},
    publisher   = {Association for Computing Machinery},
    address     = {New York, NY, USA},
    url         = {https://doi.org/10.1145/3585088.3589375},
    doi         = {10.1145/3585088.3589375},
    booktitle   = {Proceedings of the 22nd Annual ACM Interaction Design and Children Conference},
    pages       = {335–354},
    location    = {Chicago, IL, USA},
    series      = {IDC '23}
}

@article{jarvie2024age,
    author      = {Jarvie, Chelsea and Renaud, Karen},
    title       = {Online Age Verification: Government Legislation, Supplier Responsibilization, and Public Perceptions},
    journal     = {Children},
    volume      = {11},
    number      = {9},
    year        = {2024},
    doi         = {10.3390/children11091068},
    url         = {https://doi.org/10.3390/children11091068}
}

@inproceedings{yao2025age,
   author       = {Yao, Yifan and others},
   title        = {An Empirical Study on Age Verification of Adult-Oriented Apps on Google Play},
   booktitle    = {34th USENIX Security Symposium (USENIX Security 25)},
   year         = {2025},
   publisher    = {USENIX Association},
   url          = {https://www.usenix.org/system/files/usenixsecurity25-yao-yifan.pdf}
}

@misc{ofcom_age,
   author       = {{Ofcom}},
   title        = {Statement: Age Assurance and Children's Access},
   year         = {2025},
   howpublished = {\url{https://www.ofcom.org.uk/siteassets/resources/documents/consultations/category-1-10-weeks/statement-age-assurance-and-childrens-access/statement-age-assurance-and-childrens-access.pdf}}
}

@misc{ofcom_families,
   author       = {{Ofcom}},
   title        = {Families' Attitudes Towards Age Assurance},
   year         = {2022},
   howpublished = {\url{https://www.ofcom.org.uk/siteassets/resources/documents/research-and-data/online-research/keeping-children-safe-online/families-attitudes-towards-age-assurance-/drcf-ofcom-ico-age-assurance.pdf}}
}

@misc{org_uk_open_letter_2023,
   author       = {{Open Rights Group} and {European Digital Rights (EDRi)}},
   title        = {Civil society organisations urge UK to protect global digital security and safeguard private communication},
   year         = {2023},
   month        = jun,
   note         = {Open letter concerning the UK Online Safety Bill and client-side scanning},
   url          = {https://www.openrightsgroup.org/publications/open-letter-protect-encrypted-messaging/}
}

@misc{eu_csar_open_letter_2023,
   author       = {{Scientists and Researchers on the EU's Proposed Child Sexual Abuse Regulation}},
   title        = {Joint statement of scientists and researchers on EU's proposed Child Sexual Abuse Regulation},
   year         = {2023},
   month        = jul,
   note         = {Open letter addressed to Members of the European Parliament and Member States of the Council of the European Union},
   url          = {https://edri.org/wp-content/uploads/2023/07/Open-Letter-CSA-Scientific-community.pdf}
}

@misc{edri_eu_countries_no_csar_2023,
   author       = {{European Digital Rights (EDRi)} and others},
   title        = {Open letter: EU countries should say no to the CSAR mass surveillance proposal},
   year         = {2023},
   month        = sep,
   note         = {Civil society open letter to EU governments on the proposed CSA Regulation},
   url          = {https://edri.org/tag/open-letter/}
}

@misc{reuters2026_social_media_restrictions,
   author       = {{Reuters}},
   title        = {Australia, Europe countries move to curb children's social media access},
   year         = {2026},
   month        = apr,
   day          = {24},
   howpublished = {\url{https://www.reuters.com/legal/government/australia-europe-countries-move-curb-childrens-social-media-access-2026-04-24/}}
}

@online{king-schreifels2024ibelin,
  author       = {King-Schreifels, Jake},
  title        = {The Story Behind Netflix's Moving Documentary \textit{The Remarkable Life of Ibelin}},
  year         = {2024},
  date         = {2024-10-25},
  organization = {TIME},
  url          = {https://time.com/7095887/the-remarkable-life-of-ibelin-true-story-netflix/},
  urldate      = {2026-05-13}
}

@online{bbc2022basant,
  author       = {{BBC News}},
  title        = {Two Arrested in Egypt after Teenage Girl's Suicide Sparks Outrage},
  year         = {2022},
  date         = {2022-01-04},
  organization = {BBC News},
  url          = {https://www.bbc.co.uk/news/world-middle-east-59868721},
  urldate      = {2026-05-13}
}

@online{schaubert2019mats,
  author       = {Schaubert, Vicky},
  title        = {F{\o}rst da Mats var d{\o}d, forsto foreldrene verdien av gamingen hans},
  year         = {2019},
  date         = {2019-01-27},
  organization = {NRK},
  url          = {https://www.nrk.no/dokumentar/xl/forst-da-mats-var-dod_-forsto-foreldrene-verdien-av-gamingen-hans-1.14197198},
  urldate      = {2026-05-13}
}

@online{egyptianstreets2022basant,
  author       = {{Egyptian Streets}},
  title        = {Men Who Blackmailed Egyptian Girl to Suicide Get 15 Years in Jail},
  year         = {2022},
  date         = {2022-05-11},
  organization = {Egyptian Streets},
  url          = {https://egyptianstreets.com/2022/05/11/men-who-blackmailed-egyptian-girl-to-suicide-get-15-years-in-jail/},
  urldate      = {2026-05-13}
}

@misc{unesco_phone_bans_2026,
   author       = {{UNESCO Global Education Monitoring Report}},
   title        = {Phone bans in schools are spreading worldwide as the policy debate rages on},
   year         = {2026},
   month        = mar,
   day          = {19},
   howpublished = {\url{https://www.unesco.org/gem-report/en/articles/phone-bans-schools-are-spreading-worldwide-policy-debate-rages}}
}

@misc{netherlands_school_phone_rules_2025,
   author       = {{Eurydice Unit Netherlands}},
   title        = {Netherlands: A ban on mobile phones in the classroom},
   year         = {2025},
   month        = jun,
   day          = {26},
   howpublished = {\url{https://eurydice.eacea.ec.europa.eu/news/netherlands-ban-mobile-phones-classroom}}
}

@misc{france_school_phone_rules_2024,
   author       = {{Voice of America}},
   title        = {`Digital Pause': France Pilots School Mobile Phone Ban},
   year         = {2024},
   month        = sep,
   day          = {8},
   howpublished = {\url{https://www.voanews.com/a/digital-pause-france-pilots-school-mobile-phone-ban/7771105.html}}
}

@misc{chile_mobile_devices_schools_2026,
   author       = {{ECIJA}},
   title        = {Law No. 21.801: prohibition and regulation of the use of mobile devices in educational establishments},
   year         = {2026},
   month        = mar,
   day          = {2},
   howpublished = {\url{https://www.ecija.com/en/news-and-insights/ley-n-21801-prohibicion-y-regulacion-del-uso-de-dispositivos-moviles-en-establecimientos-educacionales/}}
}

@misc{greenwich_youtube_block_2025,
   author       = {Simms, Jessica},
   title        = {Greenwich Schools, Conn., Block YouTube for Elementary Students},
   year         = {2025},
   month        = mar,
   day          = {28},
   howpublished = {\url{https://www.govtech.com/education/k-12/greenwich-schools-conn-block-youtube-for-elementary-students}}
}

@misc{greenville_youtube_block_2023,
   author       = {{FOX Carolina News Staff}},
   title        = {Greenville Co. Schools blocks YouTube on devices due to potential access to inappropriate videos},
   year         = {2023},
   month        = nov,
   day          = {15},
   howpublished = {\url{https://www.foxcarolina.com/2023/11/16/greenville-co-schools-blocks-youtube-devices-due-potential-access-inappropriate-videos/}}
}

@misc{guardian2026_la_youtube_block,
   author       = {{The Guardian}},
   title        = {Los Angeles school district to limit screen time and block YouTube in classrooms},
   year         = {2026},
   month        = apr,
   day          = {22},
   howpublished = {\url{https://www.theguardian.com/us-news/2026/apr/22/los-angeles-school-district-screen-time}} 
}

@misc{roblox_parental_controls,
   author       = {{Roblox}},
   title        = {Parental Controls},
   year         = {n.d.},
   howpublished = {\url{https://about.roblox.com/parental-controls}}
}

@misc{gaggle_school_surveillance_2025,
   author       = {{Associated Press}},
   title        = {Schools use AI to monitor kids, hoping to prevent violence. Our investigation found security risks},
   year         = {2025},
   month        = mar,
   day          = {12},
   howpublished = {\url{https://apnews.com/article/25a3946727397951fd42324139aaf70f}}
}

@misc{the74_survey_goguardian_2022,
   author       = {Keierleber, Mark},
   title        = {Survey reveals extent that cops surveil students online in school and at home},
   year         = {2022},
   month        = aug,
   day          = {3},
   howpublished = {\url{https://www.the74million.org/article/survey-reveals-extent-that-cops-surveil-students-online-in-school-and-at-home/}}
}

@misc{the74_ohio_gaggle_2024,
   author       = {Keierleber, Mark},
   title        = {Ohio school districts use surveillance software to monitor student devices},
   year         = {2024},
   month        = aug,
   day          = {28},
   howpublished = {\url{https://www.the74million.org/article/ohio-school-districts-use-surveillance-software-to-monitor-student-devices/}}
}

@misc{loilo_web_filter,
   author       = {{LoiLo Inc.}},
   title        = {LoiLoNote School Pamphlet},
   year         = {2021},
   howpublished = {\url{https://assets.loilo.tv/loilonote/pdf/LNS_Pamphlets_en.pdf}}
}

@ARTICLE{rashid2025weakening,
   author       = {Rashid, Awais and May-Chahal, Corinne and Peersman, Claudia},
   journal      = {IEEE Security \& Privacy}, 
   title        = {Weakening End-to-End Encryption Considered Harmful}, 
   year         = {2025},
   volume       = {23},
   number       = {3},
   pages        = {51-54},
   doi          = {10.1109/MSEC.2025.3563092}
}

@techreport{peersman2022CSAM,
  author       = {Peersman, Claudia and De Cristofaro, Emiliano and May-Chahal, Corinne and McConville, Ryan and Rashid, Awais},
  title        = {Scoping the Evaluation of CSAM Prevention and Detection Tools in the Context of End-to-End Encryption Environments},
  institution  = {REPHRAIN},
  year         = {2022},
  month        = mar,
  number       = {Version 1.1},
  url          = {https://www.rephrain.ac.uk/wp-content/uploads/E2EE_evaluation_criteria_document24.03.21.pdf}
}

@article{volz2026age,
  title     = {Age restrictions as a tool for enhancing children’s online safety: an analysis with special reference to Australian law},
  author    = {Volz, Stephanie},
  journal   ={Computer Law \& Security Review},
  volume    ={61},
  pages     ={1--9},
  year      ={2026},
  publisher ={Elsevier}
}

@article{prendergast2026youth,
  title     = {Youth social media age restrictions: examining Trans-Tasman media coverage of Australia’s social media ‘ban’},
  author    = {Prendergast, Kate and Dyer, Emily},
  journal   = {Journal of Youth Studies},
  pages     = {1--17},
  year      = {2026},
  publisher = {Taylor \& Francis}
}

@article{archer2025coming,
  title     = {Coming of age with, in and on social media: A critique of how politicians are responding to children’s social media engagement},
  author    = {Archer, Catherine J},
  journal   = {Journal of Children and Media},
  volume    = {19},
  number    = {1},
  pages     = {58--64},
  year      = {2025},
  publisher = {Taylor \& Francis}
}

@article{adams1999users,
  title={Users are not the enemy},
  author={Adams, Anne and Sasse, Martina Angela},
  journal={Communications of the ACM},
  volume={42},
  number={12},
  pages={40--46},
  year={1999},
  publisher={ACM New York, NY, USA}
}

@inproceedings{theofanos2021passwords,
  title={'Passwords Keep Me Safe'--Understanding What Children Think about Passwords},
  author={Theofanos, Mary and Choong, Yee-Yin and Murphy, Olivia},
  booktitle={30th USENIX Security Symposium (USENIX Security 21)},
  pages={19--35},
  year={2021}
}

@article{paudel2024leveraging,
  title={Leveraging the power of storytelling to encourage and empower children towards strong passwords},
  author={Paudel, Rizu and Al-Ameen, Mahdi Nasrullah},
  journal={Proceedings of the ACM on Human-Computer Interaction},
  volume={8},
  number={CSCW2},
  pages={1--27},
  year={2024},
  publisher={ACM New York, NY, USA}
}

@inproceedings{zhang2026player,
  title={Player Safety by Design: Co-Designing Child-Centered Safety Mechanisms with Children},
  author={Zhang, Zinan and Song, Qiurong and Hernandez, Rie Helene and Liu, Yunhan and Koung, Elena and Yu, Junnan and Bai, Sunhye and Kou, Yubo and Gui, Xinning},
  booktitle={Proceedings of the 2026 CHI Conference on Human Factors in Computing Systems},
  pages={1--21},
  year={2026}
}

@inproceedings{cranor2014parents,
  title={$\{$Parents’$\}$ and $\{$Teens’$\}$ Perspectives on Privacy In a $\{$Technology-Filled$\}$ World},
  author={Cranor, Lorrie Faith and Durity, Adam L and Marsh, Abigail and Ur, Blase},
  booktitle={10th Symposium On Usable Privacy and Security (SOUPS 2014)},
  pages={19--35},
  year={2014}
}

@article{akter2022parental,
  title={From parental control to joint family oversight: Can parents and teens manage mobile online safety and privacy as equals?},
  author={Akter, Mamtaj and Godfrey, Amy J and Kropczynski, Jess and Lipford, Heather R and Wisniewski, Pamela J},
  journal={Proceedings of the ACM on Human-Computer Interaction},
  volume={6},
  number={CSCW1},
  pages={1--28},
  year={2022},
  publisher={ACM New York, NY, USA}
}

@article{feal2020angel,
  title={Angel or devil? a privacy study of mobile parental control apps},
  author={Feal, {\'A}lvaro and Calciati, Paolo and Vallina-Rodriguez, Narseo and Troncoso, Carmela and Gorla, Alessandra},
  journal={Proceedings on Privacy Enhancing Technologies},
  year={2020}
}

@inproceedings{zhao2019make,
  title={I make up a silly name' Understanding Children's Perception of Privacy Risks Online},
  author={Zhao, Jun and Wang, Ge and Dally, Carys and Slovak, Petr and Edbrooke-Childs, Julian and Van Kleek, Max and Shadbolt, Nigel},
  booktitle={Proceedings of the 2019 CHI conference on human factors in computing systems},
  pages={1--13},
  year={2019}
}

@inproceedings{ekambaranathan2021money,
  title={“Money makes the world go around”: Identifying Barriers to Better Privacy in Children’s Apps From Developers’ Perspectives},
  author={Ekambaranathan, Anirudh and Zhao, Jun and Van Kleek, Max},
  booktitle={Proceedings of the 2021 CHI conference on human factors in computing systems},
  pages={1--15},
  year={2021}
}

@inproceedings{reyes2018won,
  title={“Won’t somebody think of the children?” examining COPPA compliance at scale},
  author={Reyes, Irwin and Wijesekera, Primal and Reardon, Joel and Elazari Bar On, Amit and Razaghpanah, Abbas and Vallina-Rodriguez, Narseo and Egelman, Serge and others},
  booktitle={The 18th Privacy Enhancing Technologies Symposium (PETS 2018)},
  year={2018}
}

@inproceedings{wang202312,
  title={12 Ways to empower: Designing for children’s digital autonomy},
  author={Wang, Ge and Zhao, Jun and Van Kleek, Max and Shadbolt, Nigel},
  booktitle={Proceedings of the 2023 CHI conference on human factors in computing systems},
  pages={1--27},
  year={2023}
}

@article{wang2021protection,
  title={Protection or punishment? relating the design space of parental control apps and perceptions about them to support parenting for online safety},
  author={Wang, Ge and Zhao, Jun and Van Kleek, Max and Shadbolt, Nigel},
  journal={Proceedings of the ACM on Human-Computer Interaction},
  volume={5},
  number={CSCW2},
  pages={1--26},
  year={2021},
  publisher={ACM New York, NY, USA}
}

@article{dumaru2024s,
  title={" It's hard for him to make choices sometimes and he needs guidance": Re-orienting Parental Control for Children},
  author={Dumaru, Prakriti and Atashpanjeh, Hanieh and Al-Ameen, Mahdi Nasrullah},
  journal={Proceedings of the ACM on Human-Computer Interaction},
  volume={8},
  number={CSCW1},
  pages={1--51},
  year={2024},
  publisher={ACM New York, NY, USA}
}
\balance

\appendix

\begin{tcolorbox}[
  title=Case vignette 1,
  colback=gray!5,
  colframe=gray!60,
  fonttitle=\bfseries,
  sharp corners,
  boxrule=0.5pt
]
\textit{\textbf{Basant Khaled} was a 17-year-old Egyptian schoolgirl from Gharbia governorate. Reports indicate that she was targeted after rejecting the advances of a young man. She was reportedly sent a deceptive link, after which personal photographs from her phone were accessed and manipulated into fabricated intimate images. These images were then used to pressure and shame her, including threats that they would be shown to her family and wider community. Basant died by suicide in January 2022. Sources:~\cite{bbc2022basant,egyptianstreets2022basant}}
\end{tcolorbox}

\begin{tcolorbox}[
  title=Case vignette 2,
  colback=gray!5,
  colframe=gray!60,
  fonttitle=\bfseries,
  sharp corners,
  boxrule=0.5pt
]
\textit{\textbf{Mats Steen} was a Norwegian child with Duchenne muscular dystrophy which severely limited his offline mobility and social life. By his teenage years, much of his social interaction took place online. In World of Warcraft, he was known as Ibelin and through his online avatar, he built friendships, participated in a community, and became known not primarily through his illness but through his personality, humour, and relationships. TIME reported that he spent around 20,000 hours playing the game. After Mats died in 2014, his parents later learned that his online life had included sustained relationships, emotional support, and a strong sense of belonging. Sources:~\cite{king-schreifels2024ibelin,schaubert2019mats}}
\end{tcolorbox}

\end{document}